\documentclass[conference]{IEEEtran}
\ifCLASSINFOpdf
\else
\fi
\usepackage[utf8]{inputenc} 
\usepackage[T1]{fontenc}    
\usepackage{hyperref}       
\usepackage{url}            
\usepackage{booktabs}       
\usepackage{amsfonts}       
\usepackage{nicefrac}       
\usepackage{microtype}      
\usepackage{bm}
\usepackage{mathtools}
\usepackage{amsmath}
\usepackage{multicol}
\usepackage{multirow}
\usepackage{graphicx}
\usepackage{subcaption}
\usepackage{xcolor}
\usepackage{makecell}
\pdfoutput=1

\begin{document}
%
\title{Rethinking Recurrent Latent Variable Model for Music Composition}

\author{
    \IEEEauthorblockN{
    Eunjeong Stella Koh$^{1}$, 
    Shlomo Dubnov$^{1}$,
    Dustin Wright$^{2}$}
    
    \IEEEauthorblockA{
        \begin{tabular}{cc}
            \begin{tabular}{@{}c@{}}
                    Department of Music$^{1}$\\
                    University of California, San Diego\\
                    \{eko, sdubnov\}@ucsd.edu
            \end{tabular} & \begin{tabular}{@{}c@{}}
                    Department of Computer Science and Engineering$^{2}$\\
                    University of California, San Diego\\
                    \{dbw003\}@eng.ucsd.edu
            \end{tabular}
        \end{tabular}
    }
}

\maketitle

\begin{abstract}
We present a model for capturing musical features and creating novel sequences of music, called the Convolutional-Variational Recurrent Neural Network. To generate sequential data, the model uses an encoder-decoder architecture with latent probabilistic connections to capture the hidden structure of music. Using the sequence-to-sequence model, our generative model can exploit samples from a prior distribution and generate a longer sequence of music. We compare the performance of our proposed model with other types of Neural Networks using the criteria of Information Rate that is implemented by Variable Markov Oracle, a method that allows statistical characterization of musical information dynamics and detection of motifs in a song. Our results suggest that the proposed model has a better statistical resemblance to the musical structure of the training data, which improves the creation of new sequences of music in the style of the originals.


\end{abstract}



%
\IEEEpeerreviewmaketitle

\section{Introduction}

Neural networks have enabled automatic music composition with little human interruption. Many approaches have been proposed to generate symbolic-domain music, such as Recurrent Neural Networks (RNNs) \cite{chen2001creating, waiteproject,yang2017midinet} and RNN combined with Restricted Boltzmann Machine (RNN-RBM) \cite{boulanger2012modeling}. However, previous studies on RNN-based music generation lack in: 1) understanding the higher level semantics of a musical structure, which is critical to music composition; 2) generating novel and creative patterns that avoid literal repetitions \cite{bretan2016unit}. Most of the previous studies for music generation use so called one-to-many RNNs, where a single musical unit (such as a single note or one bar of music) is used to predict the next unit in a recurrent manner. 

In addition, recent studies exploiting Convolutional Neural Networks (CNNs) for the generation of symbolic-domain music use rich representations that are more adaptive to creating complex melodies, such as C-RNN-GAN \cite{mogren2016c}, MidiNet \cite{yang2017midinet}, and MuseGAN \cite{dong2017musegan}. In general, the frameworks' processes consist of: 1) representing multi-channel MIDI files using filters learned by CNN layers; 2) setting a discriminator to learn the distributions of melodies; and 3) processing longer sequences of data. CNNs have been well-established as choices for recognition and classification tasks in 2D data such as images, so they make better candidates for extracting melodies (horizontal) or chord (vertical) structure in musical time-pitch space.

The Variational Autoencoder (VAE) has been also explored as a generative model for creating multimedia structure. In \cite{hennig2017classifying, robertsproject}, VAE has been trained for musical creation which can better capture musical structure and generate complex sequential results. VAE exploits samples from a prior distribution and generates a longer sequence. In addition to VAE, Variational Recurrent Neural Networks have been introduced in \cite{fabius2014variational, chung2015recurrent}. These studies show that Variational Recurrent Neural Networks can create sequential data by integrating latent random variables in recurrent ways. To do this, the model utilizes encoded data in latent space in each step. This suggests that these recurrent steps can make it possible to generate more diverse styles tasks while incorporating features from data in a recognizable way. However, these previous studies do not analyze the outputs in music generation, and how to maintain a designated theme across the entire song remains unchallenged.

In this paper, we propose a Convolutional-Variational Recurrent Neural Network which combines the strength of CNN and VAE together. We show that: 1) CNN feature learning can improve statistical resemblance to musical structure of the training data; 2) utilizing encoded data in latent space can extend the dynamic creation of new sequences of music. Our model consists of a CNN to learn a better representation of bar-level of music and a Variational Recurrent Neural Network for generating novel sequences of music. In this model, random sampling and data interpolation can generate sequential data more dynamically while including learned aspects of the original structure. We model the class of bar-level data points to enable the recurrent model to infer latent variables. 

To validate our model, we adopt Information Rate (IR) as an independent measure of musical structure \cite{wang2014guided}, in order to assess the effect of the repetition versus variation structure constraints and compare our approach with that of RNN models for music generation \cite{waiteproject,yang2017midinet}. We use IR implementation by Variable Markov Oracle (VMO) to discover optimal predictive structure in the audio output of the different models. The IR analysis using VMO provides an independent evaluation of the structure of the song as captured by the sequence of audio Chroma features. Furthermore, we present a detailed motif analysis of the data and provide a qualitative discussion of generated musical samples. 

The rest of the paper is structured as follows: Section 2 gives an overview of related models and computational approaches to music generation. Section 3 describes the components involved in the Variational Recurrent Neural Network approach. Section 4 describes the IR experimental validation of the sequential modeling in the context of Nottingham dataset \cite{nottinghamdata}, a collection of 1200 British and American folk tunes. We discuss the empirical findings in Section 5 and give future perspectives.

%

\section{Backgrounds}
\subsection{Music Generation with Variational Latent Model}

Our architecture is inspired by the Variational Autoencoder (VAE) as a stochastic generative model \cite{Kingma2013AutoEncodingVB,Rezende2014StochasticBA}. In general, the model consists of a decoding network with parameters $\theta$ that estimates the posterior distribution $p_{\theta}(\mathbf{x}|\mathbf{z})$, where $\mathbf{x}$ is the sample being estimated and $\mathbf{z}$ is an unobserved continuous random variable. The prior probability $p_{\theta}(\mathbf{z})$ in this case is assumed to be generated from a Gaussian random variable with zero mean and unit variance. In this form, the true posterior distribution $p_{\theta}(\mathbf{z}|\mathbf{x}) = p_{\theta}(\mathbf{x}|\mathbf{z})p_{\theta}(\mathbf{z})/p_{\theta}(\mathbf{x})$ is intractable, so an encoding network $q$ with parameters $\phi$ is used to estimate the posterior as $q_{\phi}(\mathbf{z}|\mathbf{x})$. The encoding network is trained to estimate a multivariate Gaussian with a diagonal covariance.
\begin{equation}
\label{eq:encoderPosterior}
\log q_{\phi}(\mathbf{z}|\mathbf{x}) = \log \mathcal{N}(\mathbf{z};\bm{\mu}, \bm{\sigma}^{2}\mathbf{\text{I}})
\end{equation}
Noise can then be sampled using Gaussian distribution with the mean and standard deviation learned by the encoding network.
\begin{equation}
\label{eq:noiseSource}
\mathbf{z} = \bm{\mu} + \bm{\sigma} \odot \bm{\epsilon}, \bm{\epsilon} = \mathcal{N}(0, \mathbf{\text{I}})
\end{equation}
Thus, the parameters of the encoding network $\phi$ can be estimated with gradient descent using the re-parameterization trick \cite{Kingma2013AutoEncodingVB} and the total loss of the network is calculated as
\begin{equation}
\label{eq:vaeLoss}
\mathcal{L}(x; \theta, \phi) \simeq \frac{1}{2} \sum_{j}(1 + \log (\bm{\sigma}_{j}^{2}) -\bm{\mu}_{j}^{2} - \bm{\sigma}_{j}^{2}) + \frac{1}{L}\sum_{l} \log p_{\theta}(\mathbf{x}|\mathbf{z}^{(l)})
\end{equation}
where the first term on the right-hand side is an approximation to the KL divergence between $q_{\phi}(\mathbf{z}|\mathbf{x})$ and $p_{\theta}(\mathbf{z})$.

Intuitively speaking, the variational approach adds a probabilistic element to latent model that allows not only generation of new variations through random sampling from a noise source, but it is also trying to distill more informative latent representation by making the $\mathbf{z}$'s as independent as possible. In this view, the KL component in \autoref{eq:vaeLoss} can be considered as a probabilistic regularization that seeks the simplest or least assuming latent representation. Taking this analogy one step further, one could say that a listener infers latent variables from the musical signal she/he hears, which in turn leads her/him to imagine the next musical event by predicting musical continuation in the latent space and then "decoding" it into an actual musical sensation. 

Technically speaking, during training, the model is presented with samples of the input which are encoded by $q$ to produce the mean and standard deviation for the noise source. A noise sample is then generated and passed through the decoding network which calculates the posterior probability $p$ to determine the sample generated by the network. The network is trained to reproduce the input sample from the noise source, so the second term on the right-hand side of \autoref{eq:vaeLoss} can be either mean squared error in the case of a continuous random variable or cross entropy for discrete random variables. At test time, random samples are generated by the noise source, which is used by the decoder network to produce novel outputs.

\subsection{Music Information Dynamics and Information Rate}
We analyze our music generation output with IR value from VMO, in order to assess the predictability of a time series sequential data, and understand consistency in a song (e.g., motives, themes, etc) \cite{wang2015pattern}. VMO allows to measure music information dynamics and higher IR value presents structural note transition in a generated music than the one with lower IR value. In \autoref{eq:vmo}, $\mathbf{x}_{1},\mathbf{x}_{2},...,\mathbf{x}_{n}$ denotes time series $x$ with N observations, and $H(x)$ denotes the entropy of $x$. As a result, $IR$ denotes corresponding information between the current and previous observations, which enables the understanding of variation and repetition in a song segment. 

\begin{equation}
IR(\mathbf{x}_{1}^{n-1},\mathbf{x}_{n}) = H(\mathbf{x}_{n}) - H(\mathbf{x}_{n}|\mathbf{x}_{1}^{n-1})
\label{eq:vmo}
\end{equation}

For quantitative evaluation, VMO has also been explored by other deep learning research focusing on music generation \cite{briot2017deep,lattner2016imposing}. 

\subsection{Search for Optimal Threshold}
Evaluation of IR requires knowledge of the marginal and conditional distributions of the samples $\mathbf{x}_{n}$ and $\mathbf{x}_{1}^{n-1}$. This function is not known and the whole purpose of modeling the data with our variational latent model is to try to approximate such probabilities. So how can IR be used without an explicit knowledge of the distribution? 

The idea behind Music Information Dynamics analysis is estimating mutual information between present and past in musical data in a non-parametric way. This is done by computing similarity between features extracted from an audio signal that was synthesized from MIDI, using human engineered features and distance measures known from musical audio processing.
VMO uses a string matching algorithm, called Factor Oracle (FO), to search for repeated segments (suffixes) at every time instance in the signal. 

A crucial step in VMO is finding a threshold $\theta$ to establish similarity between features.
For each threshold value, a string compression algorithm is used to compute the mutual information between present and the past, measured in terms of the difference in the coding length of individual frames versus block encoding using repeated suffixes.
So the optimal IR in VMO is found by searching over all possible threshold values and selecting a threshold that gives an overall best compression ratio.

\begin{figure*}
\centering
\includegraphics[width=14cm, height=5cm]{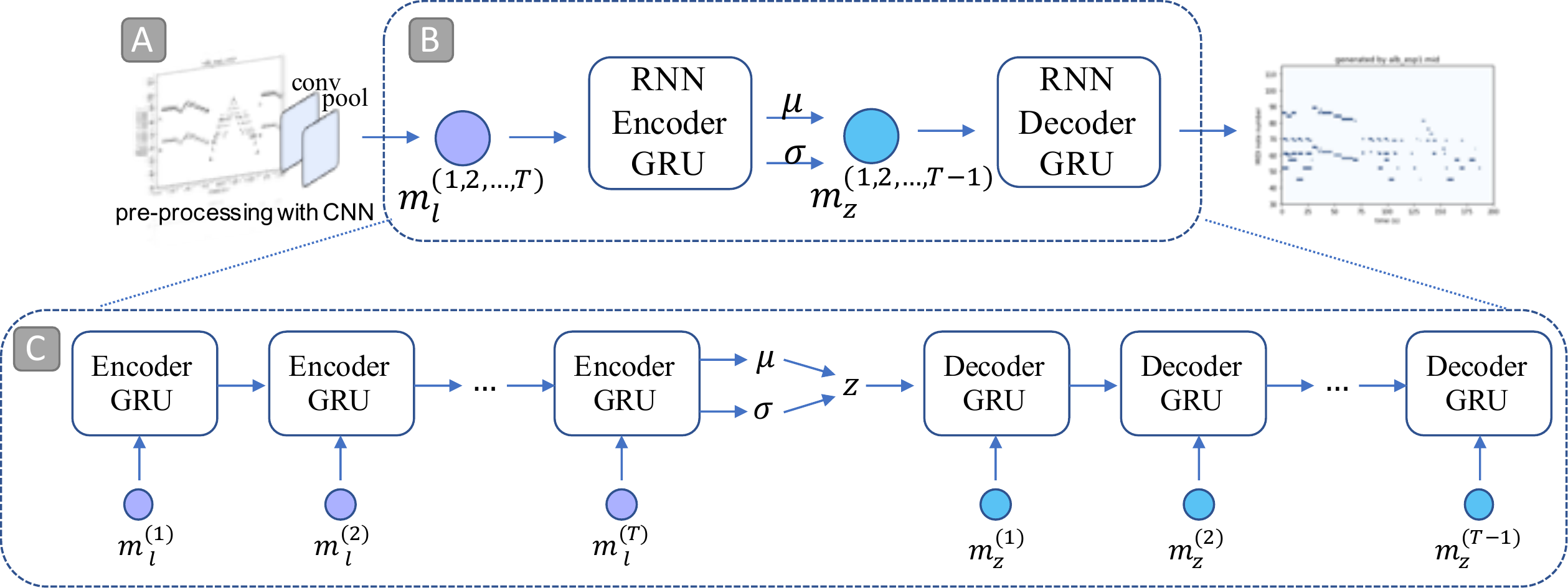} 
\label{fig:vrnn}
\caption{Convolutional-Variational Recurrent Neural Network architecture}
\label{fig:architecture}
\end{figure*}

\subsection{Links between Variational Latent Model and IR}
A motivation for using IR as a method to estimate the efficiency of dynamic latent models can be found through the relation between the variational inference loss function and IR using formulation of free energy. The loss function in \autoref{eq:vaeLoss}, known also as Evidence Lower Bound (ELBO), can be shown to represent so called free energy of the system.
\begin{equation}
\label{eq:FreeEnergy}
\mathcal{L} \simeq E_{q} [\log p(\mathbf{x},\mathbf{z}) -
\log q(\mathbf{z}|\mathbf{x})] = -\mathcal{F}
\end{equation}
Let us further assume that the samples $x$ depend only on the most recent $z$. In such case, the first term averaged by $q$ over all possible $z$ values approximately corresponds to negative of marginal entropy of the data $x$, $-H(x_n)$. The second term  captures the entropy of $z$ that contains the residual information in the measurements, similar to information that is captured by the entropy rate of $x$ as $H(x_n | x_1^{n-1})$ for asymptotically large $n$. Under such assumptions $-\mathcal{F}$ is similar to the IR expression given in \autoref{eq:vmo}. 
Accordingly, finding the minimum of $\mathcal{L}$ is equivalent to maximizing $\mathcal{F}$, which in case of our time signal assumptions\footnote{It is worth noting that we are assuming here that the entropy of the latent states is equal to entropy rate of the data.} approximately equals to IR.

\section{methods}
\subsection{Feature Learning with CNN}

We adopt a CNN in order to learn a better representation of polyphonic music by treating the input as a 2D binary feature map. This is predicated on the notion that the arrangement of notes in a musical piece yields salient spatial relationships when visualized in a form such as a piano-roll and thus are conducive to being modeled by a CNN. In this, the input MIDI is first preprocessed into a piano-roll, with the beat resolution set to 8\textsuperscript{th} notes. This gives us a feature map representation $\mathbf{x}^{(t)} \in \{0,1\}^{n \times r \times 1}$ at time step $t$, where $n$ is a number of time steps in a frame and $r$ is the note range. The piano-roll is then processed by a CNN with two convolutional layers separated by max-pooling layers and a final flattening layer. The output of this network is the latent feature $\mathbf{m}_{l}^{(t)} \in \mathbb{R}^{k}$ at time step $t$ (See Panel A in \autoref{fig:architecture}).

\subsection{Encoder \& Decoder Network}
The models presented in \autoref{fig:architecture} can generate a track of music bar by bar, with a possibly polyphonic structure among several bars. We adopt a recurrent architecture for our VAE, which includes an RNN encoder and RNN decoder (See Panel B and C). The encoder RNN takes the latent feature $\mathbf{m}_{l}^{(t)}$ at each time step and produces a final hidden state $\mathbf{h}_{q}^{(T)} \in \mathbb{R}^{e}$ for a sequence of $T$ MIDI frames. 
\begin{equation}
\mathbf{h}_{q}^{(T)} = f_{\text{RNN}}(\mathbf{m}_{l}^{(1)}, ..., \mathbf{m}_{l}^{(T)})
\end{equation}
The hidden state is then subject to two linear transformations to determine the mean and standard deviation of the noise distribution given in \autoref{eq:encoderPosterior}.
\begin{equation}
\begin{split}
\bm{\mu} = \mathbf{W}_{\mu}\mathbf{h}_{q}^{(T)} + \mathbf{b}_{\mu} \\
\bm{\sigma} = \mathbf{W}_{\sigma} \mathbf{h}_{q}^{(T)} + \mathbf{b}_{\sigma}
\end{split}
\end{equation}
Where $\mathbf{W}_{\mu}, \mathbf{W}_{\sigma} \in \mathbb{R}^{z \times e}$ and $\mathbf{b}_{\mu}, \mathbf{b}_{\sigma} \in \mathbb{R}^{z}$. Noise is then generated as in \autoref{eq:noiseSource}.
Since we are modeling sequential data, the decoder network is trained to predict $p_{\theta}(\mathbf{x}^{(t)} | \mathbf{x}^{(1:t-1)}, \mathbf{z})$. In this, the RNN takes in the generated noise $\mathbf{z}$ at the first time step. At each subsequent time step, the latent feature $\mathbf{m}_{l}^{(t)}$ for an input sample $\mathbf{x}^{(t)}$ is linearly transformed into the same dimensionality as the noise source and then passed into the RNN.
\begin{equation}
\mathbf{m}_{z}^{(t)} = \mathbf{W}_{z}\mathbf{m}_{l}^{(t)} + \mathbf{b}_{z}
\end{equation}
Where $\mathbf{W}_{z} \in \mathbb{R}^{z \times k}$ and $\mathbf{b}_{z} \in \mathbb{R}^{z}$. The RNN produces a hidden state $\mathbf{h}_{p}^{(t)}$ at each time step, which is passed through a logistic layer estimating $p_{\theta}(\mathbf{x}^{(t)} | \mathbf{x}^{(1:t-1)}, \mathbf{z})$.
\begin{equation}
\mathbf{h}_{p}^{(t)} = f_{\text{RNN}}(\mathbf{z}, \mathbf{m}_{z}^{(1)}, ..., \mathbf{m}_{z}^{(t)})
\end{equation}
\begin{equation}
\label{eq:logistic}
\tilde{\mathbf{x}}^{(t)} = \sigma(\mathbf{W}_{p}\mathbf{h}_{p}^{(t)} + \mathbf{b}_{p})
\end{equation}
Where $\sigma(\cdot)$ is the logistic sigmoid function, $\mathbf{h}_{p}^{(t)} \in \mathbb{R}^{d}$, and $\mathbf{W}_{p} \in \mathbb{R}^{nr \times d}$. This effectively yields a binary feature map of the same dimensionality as the input which is used to predict a piano-roll based on the input at the previous time steps and the noise. Finally, we use the Gated Recurrent Unit (GRU) \cite{Cho2014LearningPR} for both the encoder and decoder RNN, which is defined by the following equations for $f_{\text{RNN}}(\mathbf{x}^{(1)}, ..., \mathbf{x}^{(t)})$ at time step $t$.

\begin{equation*}
\mathbf{s}^{(t)} = \sigma_{g}(\mathbf{W}_{s}\mathbf{x}^{(t)} + \mathbf{U}_{s}\mathbf{h}^{(t-1)} + \mathbf{b}_{s})
\end{equation*}
\begin{equation*}
\mathbf{r}^{(t)} = \sigma_{g}(\mathbf{W}_{r}\mathbf{x}^{(t)} + \mathbf{U}_{r}\mathbf{h}^{(t-1)} + \mathbf{b}_{r})
\end{equation*}
\begingroup
\small
\begin{equation*}
\mathbf{h}^{(t)} = \mathbf{s}^{(t)} \odot \mathbf{h}^{(t-1)} + (1 - \mathbf{s}^{(t)}) \odot \sigma_{h}(\mathbf{W}_{h}\mathbf{x}^{(t)} + \mathbf{U}_{h}(\mathbf{r}^{(t)} \odot \mathbf{h}^{(t-1)}) + \mathbf{b}_{r})
\end{equation*}
\endgroup

Here, $\sigma_{g}(\cdot)$ denotes the logistic sigmoid function and $\sigma_{h}(\cdot)$ denotes hyperbolic tangent. In our implementation, we use 256 hidden units for the encoder and 512 hidden units for the decoder. With GRU, the model can create sequential output combined with decoded noise and previous output utilized for next input. As shown in \autoref{fig:architecture}, the model sequentially generates bars one after another based on VAE structure, which takes inputs of mean and variance, then proceeding to next step which processes a random noise z and output received by the previous GRU. 

\subsection {Training Details}
We train the network on training MIDI files, segmenting the MIDI input into batches of 8 bars, half a bar to a time step, and 8th note as the note resolution, resulting in 16 time steps. In each epoch, we train on the entire song with non-overlapping batches. We use dropout as a regularizer on the output of the CNN and the output of the decoder RNN. For optimization, we use the Adam optimizer with a learning rate of 0.001. In addition, we clip the gradients of the weight matrices so the L2 norms are less than 10. The loss of our network is that of \autoref{eq:vaeLoss}, with the log loss in the second term being cross entropy loss between the input samples and the output of the decoder. The model generally converges around 200 epochs. By enabling loss function calculation automatically, we observe and measure the model by the cost function. 
During training, our model is to focus the posterior of probability by training network to process the mean and variance of this posterior. In the aspects of variational inference, as the learning is repeated, the difference is minimized (\autoref{eq:vaeLoss}). 

\section{Experiments}
To evaluate the structural quality of the musical result, we compare our model with the MelodyRNN model \cite{waiteproject}. The MelodyRNN model is designed in several different ways (basic RNN, lookback RNN, attention RNN, and polyphony RNN), and we chose attention RNN and polyphony RNN model, which allow the model to capture longer dependencies, and result in melodies that involve arching themes \cite{Melodyproject}. Specifically, polyphony RNN aims at polyphonic music generation, so it is an appropriate baseline to compare with our model. For our experiments, our training data comes from the Nottingham Dataset, a collection of 1200 folk songs \cite{nottinghamdata}. Each training song is segmented into frames (piano-roll), and for the preprocessing of our dataset, we implement our method based on the \texttt{music21,librosa}, and \texttt{pretty\_midi} packages for feature extraction on MIDI file \cite{cuthbert2010music21, mcfee2015librosa, raffeldata}. We use an input of 128 binary visible units and aligned on the 8th note beat level. With these data, we train each model of MelodyRNN and our proposed network to create MIDI sequences. Both our proposed model and MelodyRNN model converge around 200 epochs. Our implementation is now available on github\footnote{https://github.com/skokoh/c\_vrnn\_mmsp\_2018}.

\begin{table}
\begin{center}
\begin{tabular}{  c | c | c | c  }

\hline
Melodies & \makecell{Total IR\\(8 bars)} & \makecell{Total IR\\(16 bars)} & \makecell{Total IR\\(32 bars)}\\
\hline
\hline

Nottingham Original \cite{nottinghamdata} & 4974.61 & 7412.91 & 18567.01 \\ 
\hline
Proposed & 3463.81 & 6047.28 & 16044.91 \\
\hline

PolyphonyRNN \cite{waiteproject} & 3023.44 & 6027.04 & 15425.27 \\ 
\hline
AttentionRNN \cite{waiteproject} & 3381.71 & 5712.87 & 14192.60 \\ 
\hline

MidiNet \cite{yang2017midinet} & 3117.68 & - & - \\
\hline\hline
Time(s) & 15.3 & 29.2 & 67 \\
\hline

\end{tabular}
\end{center}
\caption{Total IR Results (Averaged scores)}
\label{tab:cnn}
\end{table}

\subsection{Model comparison}
After training, we compare generated samples from each of 3 models (proposed, polyphony RNN, attention RNN) for each of 3 settings in generated sample duration of 8 bars, 16 bars, and 32 bars. We use IR from VMO \cite{wang2015pattern} as a basis of comparison and each generated MIDI sample was synthesized to audio signal. For comparison, we extracted 30 unique generated songs from each setting (See \autoref{tab:cnn}), thus 273 individual sample songs are tested for evaluation\footnote{3 audio samples (8 bars length only) are generated by MidiNet Model, which are uploaded on \url{https://github.com/RichardYang40148/MidiNet/tree/master/v1/}}. In the case of MidiNet, only 3 different testing samples are available within 8 bar length of audio sample. We want to see the variation in Total IR value which could be affected by the length of song in structural analysis. We report an averaged value of IR in \autoref{tab:cnn}. 

We empirically analyze our model in several settings against MelodyRNN model. We share key observations:

$\bullet$ \autoref{tab:cnn} shows average IRs for original Nottingham MIDI datasets and for generated samples from several models, where higher IRs report more distinct self-similarity structures. The IR of the original dataset is higher than that of the generated music. Self-similarity in audio refers to the multi-scalar feature in a set of relationships, and it commonly indicates musical coherence and consistency \cite{foote1999visualizing}. 

$\bullet$ In \autoref{tab:cnn}, polyphony RNN and attention RNN models present lower IR than our proposed model does. Results in each setting show that the convolutional recurrent latent variable sampling approach increases the IR of the produced musical material over other neural network approaches, indicating a higher degree of structure. Accordingly, the results manifest our proposed model can generate higher level of musically consistency structure.

\begin{figure}[h]
\centering

\includegraphics[width=\linewidth]
{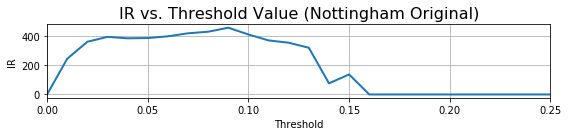}

\includegraphics[width=\linewidth]
{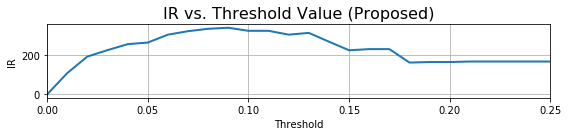}


\includegraphics[width=\linewidth]
{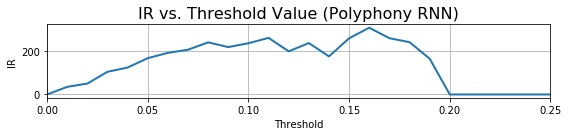}

\includegraphics[width=\linewidth]
{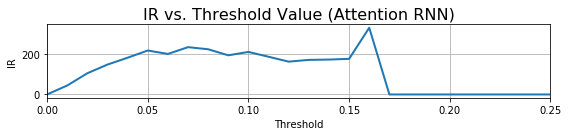}

  \caption{Total IR vs. Threshold $\theta$ value (VMO)}
  \label{fig:threshold}
\end{figure}

$\bullet$ In \autoref{fig:threshold}, the visualizations of the IR values versus different $\theta$ on one song are represented. From top to bottom, we share the results of the sample songs from each setting, training dataset, our proposed model, polyphony RNN, and attention RNN model. The results show the relation between IR and threshold value and implies different musical structures are generated by different $\theta$ values. In terms of the results graphs, the attention RNN recopies longer segments, but they are interrupted, which is dropped down in figure, while our proposed model relies on shorter previous patterns, but the transitions are smoother thus the blocks are longer. 

\begin{figure}[h]
\centering

\includegraphics[width=\linewidth]
{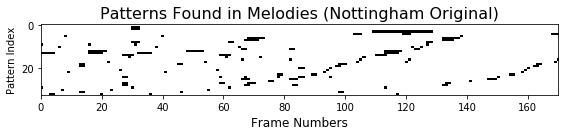}

\includegraphics[width=\linewidth]
{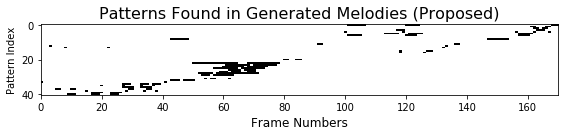}

\includegraphics[width=\linewidth]
{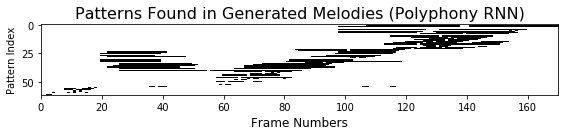}

\includegraphics[width=\linewidth]
{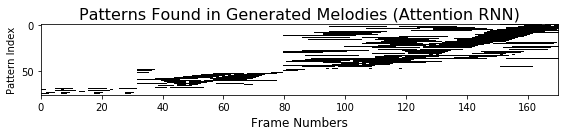}

  \caption{Pattern Findings with VMO}
  \label{fig:patterns}
\end{figure}

$\bullet$ In \autoref{fig:patterns}, the results for finding repeated patterns in one of the audio samples generated from each setting are displayed from top to bottom. The y-axis indicates the pattern index of repeated motifs of a signal sampled at discrete times shown along the x-axis. The lines represent repeated motifs, which are longer and fewer in the RNN case. In the graph from Nottingham Original, we can recognize that the original has many more shorter musical pattern indexes appearing at multiple frame numbers. The overall distribution of repeated themes seems to be captured better in the outputs of our proposed approach, suggesting that it captures some structural aspects of patterns distribution of the data as well.




\subsection{Application}

In this section, we share our generated melodies in terms of following the research question: can we build a model capable of learning long-term structure and capable of including the method to generate polyphonic music pieces? 

\begin{figure}[h]
\centering
\includegraphics[width=\linewidth]
{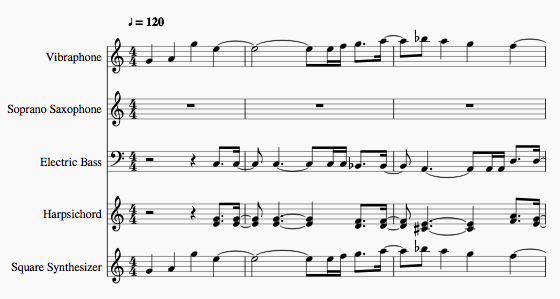}

\vspace*{+2mm}
\includegraphics[width=\linewidth]
{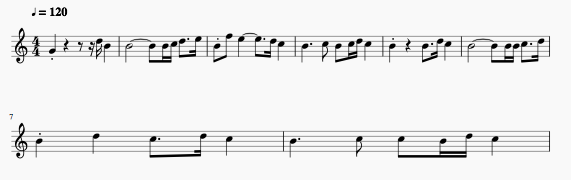}

\vspace*{+2mm}
\includegraphics[width=\linewidth]
{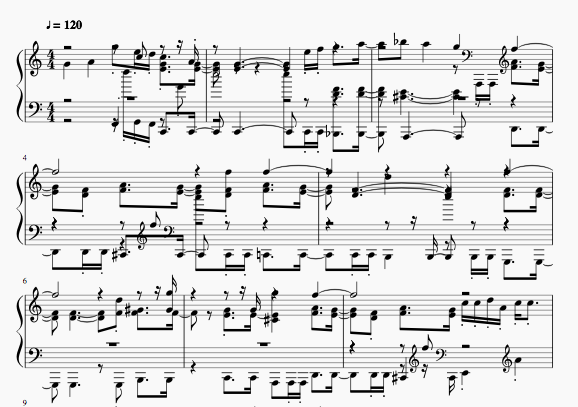}
  \caption{Examples from the Video Game music sample and generated results from attention RNN and proposed method. From top to bottom: original sonic\_starlight\_zone.mid, attention RNN result, proposed method result.}
  \label{fig:generated_sonic}
\end{figure}

Considering an application level, we explore video game music generation and emulate a specific song from music samples for creating a new sequence of music (See \autoref{fig:generated_sonic}). By doing this, we use 10 different MIDI files derived from a corpus of Video Game music\footnote{\url{https://www.vgmusic.com}} and we generate 10 unique MIDI outputs based on each training sample. The MIDI files are mainly composed of 4-5 different instruments with multi-tracks. From this approach, our model copies the theme from previous music sample and mimics the style of music with a new sequence. 


In \autoref{fig:generated_sonic}, the results indicate that our proposed model can generate music beyond monophonic melodies for various types of music, depending on the input data. The result of the attention RNN differs in our model and in the complexity of the results, since attention RNN model covers simple melody generation/progression and repetitive patterns appearing in the generation results. Our generated melodies shows that we can create long-term structure of music and can compose complex sequence of music while including the original theme. Moreover, our proposed model can process training samples from a prior distribution and generate the sequence more dynamically. Our sample results for video game music are also posted on soundcloud\footnote{https://soundcloud.com/user-431911640/sets}. 


\section {Discussion}
In this study, we show initial proof that our proposed model applied to MIDI sequence representations can capture the structure of the song and create polyphonic music. The motivations behind combining CNN, RNN and VAE were to explore significant problems in music generation which are related to representation issues that are handled via CNN, repetitive patterns in generated output that are known in RNN and ability to generate variations from progression of melody sequence. In our study, we used IR as a critera to evaluate the generated output and compare it to other models. 

From the quantitative evaluation, the results show that the latent variable sampling approach substantially increases the IR of the generated musical material over other neural network approaches, implying a higher degree of semantic structure. At the application stage of our method, we introduce the model to emulate a specific song from a video game and generate background music similar in style to those examples. Some musical applications need to work with fewer samples in order to generate a specific musical result and our Convolutional-Variational Recurrent Neural Network would be flexible about the size of dataset. 

In addition to VAE utilized in this paper, other generative models have been actively challenged in different ways for music generation purpose. Given the recent enthusiasm in deep learning with music, we also practice introducing combined neural network models and data representations that effectively process the melodic polyphonic harmonic structure in music.

\section*{Acknowledgment} 

The authors would like to thank Richard Yang, who generously shared one of their training data from MidiNet\cite{yang2017midinet}. 



%



\bibliographystyle{IEEEtran}
\bibliography{references}

\begin{thebibliography}{10}
\providecommand{\url}[1]{#1}
\csname url@samestyle\endcsname
\providecommand{\newblock}{\relax}
\providecommand{\bibinfo}[2]{#2}
\providecommand{\BIBentrySTDinterwordspacing}{\spaceskip=0pt\relax}
\providecommand{\BIBentryALTinterwordstretchfactor}{4}
\providecommand{\BIBentryALTinterwordspacing}{\spaceskip=\fontdimen2\font plus
\BIBentryALTinterwordstretchfactor\fontdimen3\font minus
  \fontdimen4\font\relax}
\providecommand{\BIBforeignlanguage}[2]{{%
\expandafter\ifx\csname l@#1\endcsname\relax
\typeout{** WARNING: IEEEtran.bst: No hyphenation pattern has been}%
\typeout{** loaded for the language `#1'. Using the pattern for}%
\typeout{** the default language instead.}%
\else
\language=\csname l@#1\endcsname
\fi
#2}}
\providecommand{\BIBdecl}{\relax}
\BIBdecl

\bibitem{chen2001creating}
C.-C. Chen and R.~Miikkulainen, ``Creating melodies with evolving recurrent
  neural networks,'' in \emph{Neural Networks, 2001. Proceedings. IJCNN'01.
  International Joint Conference on}, vol.~3.\hskip 1em plus 0.5em minus
  0.4em\relax IEEE, 2001, pp. 2241--2246.

\bibitem{waiteproject}
E.~Waite, D.~Eck, A.~Roberts, and D.~Abolafia, ``Project magenta,'' available
  at
  \url{https://github.com/tensorflow/magenta/tree/master/magenta/models/melody_rnn}.

\bibitem{yang2017midinet}
L.-C. Yang, S.-Y. Chou, and Y.-H. Yang, ``Midinet: A convolutional generative
  adversarial network for symbolic-domain music generation,'' in
  \emph{Proceedings of the 18th International Society for Music Information
  Retrieval Conference (ISMIR’2017), Suzhou, China}, 2017.

\bibitem{boulanger2012modeling}
N.~Boulanger-Lewandowski, Y.~Bengio, and P.~Vincent, ``Modeling temporal
  dependencies in high-dimensional sequences: application to polyphonic music
  generation and transcription,'' in \emph{Proceedings of the 29th
  International Coference on International Conference on Machine
  Learning}.\hskip 1em plus 0.5em minus 0.4em\relax Omnipress, 2012, pp.
  1881--1888.

\bibitem{bretan2016unit}
M.~Bretan, G.~Weinberg, and L.~Heck, ``A unit selection methodology for music
  generation using deep neural networks,'' \emph{arXiv preprint
  arXiv:1612.03789}, 2016.

\bibitem{mogren2016c}
O.~Mogren, ``C-rnn-gan: Continuous recurrent neural networks with adversarial
  training,'' \emph{arXiv preprint arXiv:1611.09904}, 2016.

\bibitem{dong2017musegan}
H.-W. Dong, W.-Y. Hsiao, L.-C. Yang, and Y.-H. Yang, ``Musegan: Symbolic-domain
  music generation and accompaniment with multi-track sequential generative
  adversarial networks,'' \emph{arXiv preprint arXiv:1709.06298}, 2017.

\bibitem{hennig2017classifying}
J.~A. Hennig, A.~Umakantha, and R.~C. Williamson, ``A classifying variational
  autoencoder with application to polyphonic music generation,'' \emph{arXiv
  preprint arXiv:1711.07050}, 2017.

\bibitem{robertsproject}
A.~Roberts, J.~Engel, and D.~Eck, ``Musicvae,'' available at
  \url{https://github.com/tensorflow/magenta/tree/master/magenta/models/music_vae}.

\bibitem{fabius2014variational}
O.~Fabius and J.~R. van Amersfoort, ``Variational recurrent auto-encoders,''
  \emph{arXiv preprint arXiv:1412.6581}, 2014.

\bibitem{chung2015recurrent}
J.~Chung, K.~Kastner, L.~Dinh, K.~Goel, A.~C. Courville, and Y.~Bengio, ``A
  recurrent latent variable model for sequential data,'' in \emph{Advances in
  neural information processing systems}, 2015, pp. 2980--2988.

\bibitem{wang2014guided}
C.-i. Wang and S.~Dubnov, ``Guided music synthesis with variable markov
  oracle,'' in \emph{The 3rd international workshop on musical metacreation,
  10th Artificial intelligence and interactive digital entertainment
  conference}, 2014.

\bibitem{nottinghamdata}
``Nottingham database,'' available at
  \url{http://ifdo.ca/~seymour/nottingham/nottingham.html}.

\bibitem{Kingma2013AutoEncodingVB}
D.~P. Kingma and M.~Welling, ``Auto-encoding variational bayes,'' \emph{CoRR},
  vol. abs/1312.6114, 2013.

\bibitem{Rezende2014StochasticBA}
D.~J. Rezende, S.~Mohamed, and D.~Wierstra, ``Stochastic backpropagation and
  approximate inference in deep generative models,'' in \emph{ICML}, 2014.

\bibitem{wang2015pattern}
C.-i. Wang and S.~Dubnov, ``Pattern discovery from audio recordings by variable
  markov oracle: A music information dynamics approach,'' in \emph{Acoustics,
  Speech and Signal Processing (ICASSP), 2015 IEEE International Conference
  on}.\hskip 1em plus 0.5em minus 0.4em\relax IEEE, 2015, pp. 683--687.

\bibitem{briot2017deep}
J.-P. Briot, G.~Hadjeres, and F.~Pachet, ``Deep learning techniques for music
  generation-a survey,'' \emph{arXiv preprint arXiv:1709.01620}, 2017.

\bibitem{lattner2016imposing}
S.~Lattner, M.~Grachten, and G.~Widmer, ``Imposing higher-level structure in
  polyphonic music generation using convolutional restricted boltzmann machines
  and constraints,'' \emph{arXiv preprint arXiv:1612.04742}, 2016.

\bibitem{Cho2014LearningPR}
K.~Cho, B.~van Merrienboer, Çaglar Gulçehre, D.~Bahdanau, F.~Bougares,
  H.~Schwenk, and Y.~Bengio, ``Learning phrase representations using rnn
  encoder-decoder for statistical machine translation,'' in \emph{EMNLP}, 2014.

\bibitem{Melodyproject}
``Melodyrnn,'' available at
  \url{https://github.com/tensorflow/magenta/tree/master/magenta/models/}.

\bibitem{cuthbert2010music21}
M.~S. Cuthbert and C.~Ariza, ``music21: A toolkit for computer-aided musicology
  and symbolic music data,'' 2010.

\bibitem{mcfee2015librosa}
B.~McFee, C.~Raffel, D.~Liang, D.~P. Ellis, M.~McVicar, E.~Battenberg, and
  O.~Nieto, ``librosa: Audio and music signal analysis in python,'' in
  \emph{Proceedings of the 14th python in science conference}, 2015, pp.
  18--25.

\bibitem{raffeldata}
C.~Raffel and D.~P. Ellis, ``Data with pretty\_midi.''

\bibitem{foote1999visualizing}
J.~Foote, ``Visualizing music and audio using self-similarity,'' in
  \emph{Proceedings of the seventh ACM international conference on Multimedia
  (Part 1)}.\hskip 1em plus 0.5em minus 0.4em\relax ACM, 1999, pp. 77--80.

\end{thebibliography}





\end{document}